\documentclass[runningheads]{llncs}
\usepackage[T1]{fontenc}
\usepackage{graphicx}
\usepackage{placeins}
\usepackage{float}
\usepackage{amsmath}
\usepackage{amssymb}
\usepackage{makecell}
\usepackage{gensymb}
\usepackage{subcaption}
\usepackage{xurl}

\begin{document}
\title{Towards an End-to-End System for \\3D Tracking of Physical Objects \\in Virtual Immersive Environments}
\titlerunning{Towards a System for 3D Tracking in VIE}
%
\author{Stanisław Knapiński\inst{1}\orcidID{0009-0006-8861-3136} \and
Maciej Grzeszczuk\inst{1}\orcidID{0000-0002-9840-3398} \and
Barbara Karpowicz\inst{1}\orcidID{0000-0002-7478-7374} \and
Pavlo Zinevych\inst{1}\orcidID{0009-0008-9250-8712} \and
Wieslaw Kopec\inst{1}\orcidID{0000-0001-9132-4171}} 
%
\authorrunning{S. Knapiński et al.}
%
\institute{Polsko-Japońska Akademia Technik Komputerowych w Warszawie,\\ Koszykowa 86, Warszawa, Polska\\}
\maketitle              
\begin{abstract}
This work aims to establish an end-to-end system for tracking of physical 3D objects for virtual reality (VR) applications. We focus on training applications requiring real-time tracking of the position of small physical objects and their reflection in VR space. Out goal is to perform object tracking in a "plug and play" manner, without using complex systems with quite large tracking devices or manually implementing object tracking. We therefore propose a system for object tracking via fiducial markers alongside a software harness, to enable fast and efficient designation of objects to be tracked and data streaming solution for end-use applications. The system utilizes AruCo \cite{AruCoMarkers}, AprilTag \cite{AprilTag} and an original Colored Control Points based fiducial system. It allows for easy tag detection and use of object position data, which are crucial for immersive training environments based on VR and eXtended Reality (XR). We evaluate various tag sizes, detection distances, and different camera devices against the theoretical limits. In effect, we create a complete solution for implementing marker-based, real-to-virtual object position mapping for various applications.

\keywords{3D Object Tracking \and Virtual Reality \and eXtended Reality \and Fiducial Markers}
\end{abstract}
%
%
%
\section{Introduction}
The MR-Continuum\footnote{Reality-Virtuality continuum describes all technologies ranging from fully virtual reality to augmented reality, and everything in between} market has been rapidly growing in recent years \cite{iburahim2025virtual}. This technology field inadvertently requires solving of intricate challenges, one of them being proper and valid matching of the real world with the virtual one in real time \cite{slater2014grand,SRINIVASAN1997393}. Particularly, the challenge of precisely matching locations of real objects to objects in virtual space, is what motivates the entirety of this work.

Currently, the most used and developed way of tracking objects in 3D space in virtual/extended reality applications are VR Trackers \cite{ViveTracker}. Some systems exist to allow for tracking motion without any extra hardware \cite{kapsoritakis2022comparative}, but the application is limited to hands and eyes only. The robotics field has developed other ways to track objects, such as the AprilTag \cite{AprilTag}, but these are not widely used in the VR/XR world and usually limited to the user's field of view (e.g. Varjo \cite{Varjo} systems). Any other objects which we would like to track independently of the user's field of view in virtual or mixed environment would currently need the tracker, and therefore provides a vital limitation for systems that require manipulation of relatively small objects (such as environments used for training).

\section{Related Works}
\begin{figure}[htbp!]
    \centering
    \includegraphics[width=\textwidth]{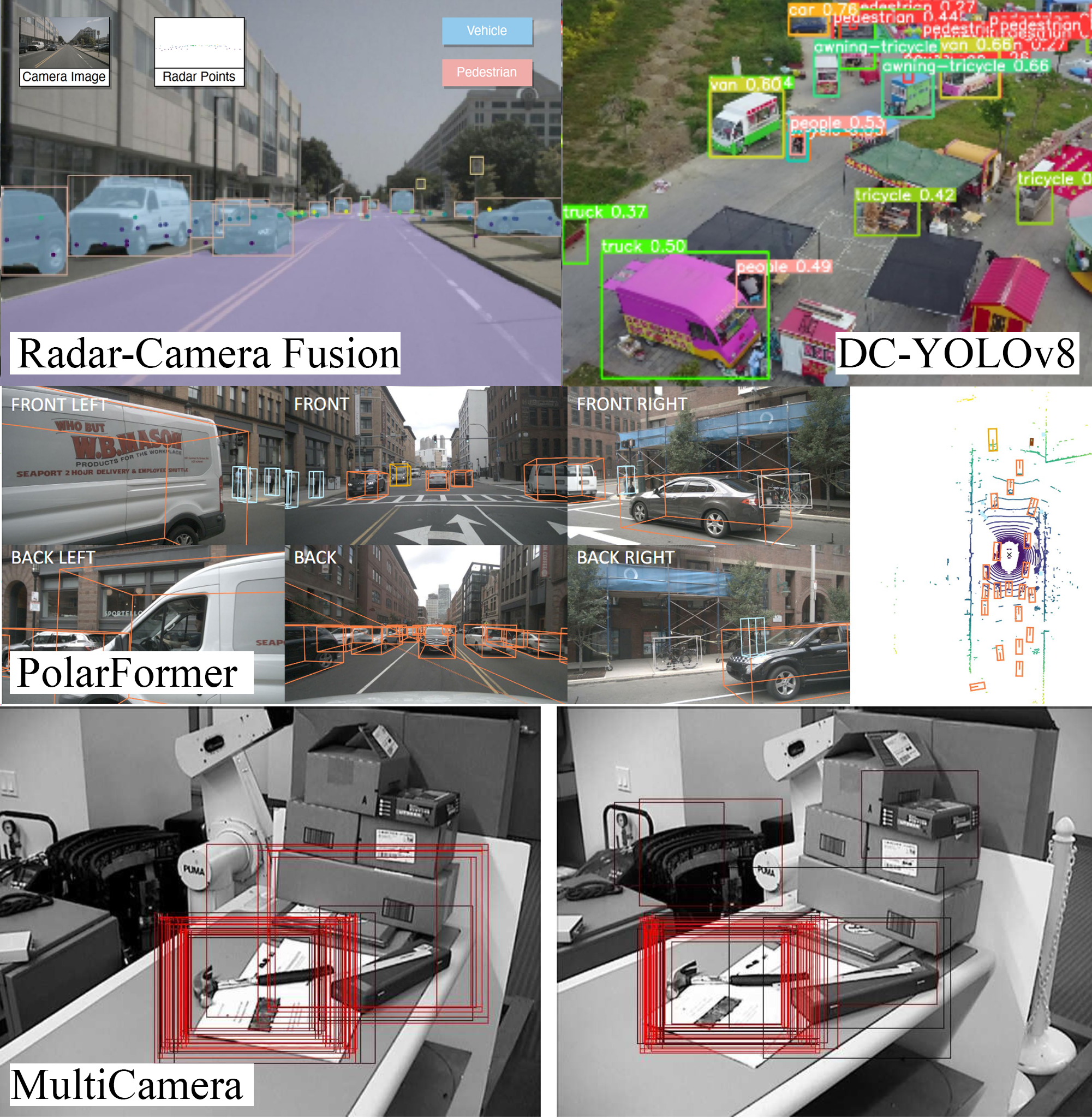}
    \caption{Object detection systems. Source: Own elaboration, \cite{yao2023radar,jiang2023polarformer,coates2010multi,lou2023dc}}
    \label{fig:obj_det_sys}
\end{figure}

Taking a frame of video or a still image, distinguishing various objects of interest and then extracting data about them is collectively known as \textbf{object detection}. Various systems (see Figure \ref{fig:obj_det_sys}) such as LiDAR, 2D image-based, radar and other exist, performing objects detection and classification. Those include Radar-Camera Fusion \cite{yao2023radar}, Polarformer \cite{jiang2023polarformer}, Multi-Camera \cite{coates2010multi} and DC-YOLOv8 \cite{lou2023dc}.

While the robotics field uses cameras and computer vision algorithms to distinguish objects, in the VR world, the most common approach \cite{SteamVRTracking} is to use sensors to estimate position and motion of objects. The primary concern is usually motion tracking, to allow proper human-computer interaction. VR tracking solutions include Valve's Lighthouse-based System \cite{SteamVRTracking} with open hardware alternatives \cite{ng2017low}, Metaspace II \cite{sra2015metaspace} and GBOT \cite{li2024gbot}.

The space for improvement is visible in training applications, where real-time tracking of small physical objects and their accurate reflection in virtual space is required. The use of fiducial markers \cite{skromme2015step} should allow efficient tracking of physical objects and direct data streaming to applications like real-time rendering engines.

\section{Proposed Solution and Implementation}
In our opinion, tracking technology should not depend on active tracking devices, due to their large size. Fiducial markers are small identifiers attached to objects, that allow the software to distinguish them using a simple camera, and through the use of pose estimation, find their angles and position. In our work, we use AprilTags \cite{AprilTag}, AruCo Markers \cite{AruCoMarkers} and Colored Points method for the task.

\begin{figure}[htbp!]
    \centering
    \includegraphics[width=0.6\textwidth]{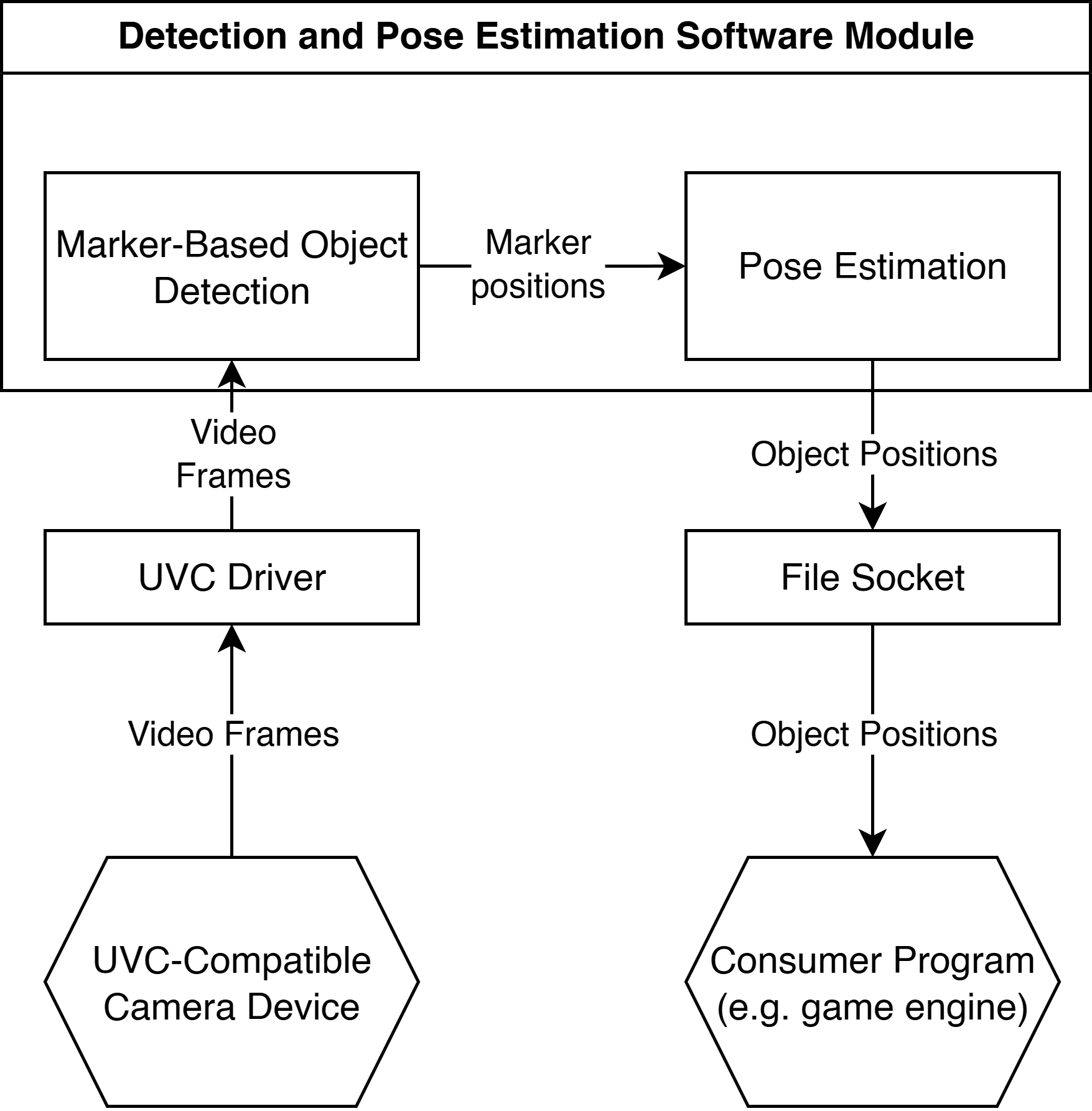}
    \caption{Example workflow with the object detection software. Source: own elaboration}
    \label{fig:modstruct_objdet_sys}
\end{figure}

\begin{figure}[htbp!]
    \centering
    \includegraphics[width=\linewidth]{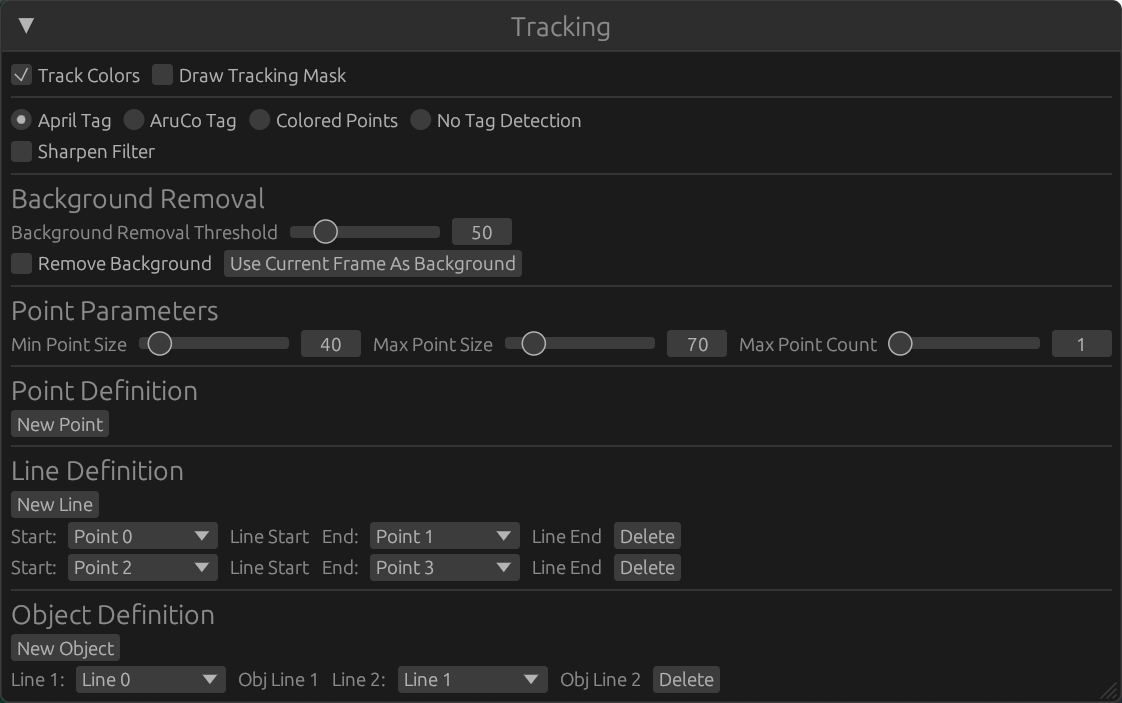}
    \caption{Tracking Configuration UI. Source: own elaboration}
    \label{fig:trackcfgui}
\end{figure}

We propose a software solution (see Figure \ref{fig:modstruct_objdet_sys}) developed with Rust \cite{Rust}, due to its memory safety and speed guarantees \cite{bugden2022rustprogramminglanguagesafety}, that calibrates the camera, allows the user to set up (shown in Figure \ref{fig:trackcfgui}) the tracking method easily, and feed the object coordinates to other software, such as the Unity \cite{UnityEngine} game engine. A graphical user interface, designed with the \textit{egui} library \cite{Egui} allows setup and monitoring.

\subsection{Pose Estimation and Tracking}
\begin{figure}[htbp]
\centering
\begin{minipage}{.48\textwidth}
  \centering
  \includegraphics[width=\linewidth]{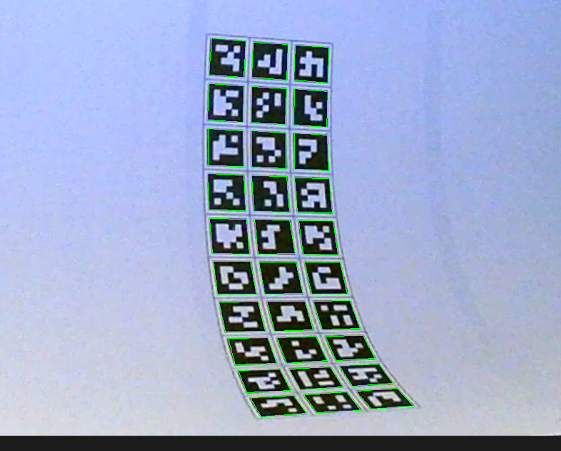}
  \captionof{figure}{A sheet of AruCo markers being detected. Source: own elaboration.}
  \label{fig:arcuosheet}
\end{minipage}%
\begin{minipage}{0.04\textwidth}
\includegraphics[width=\linewidth]{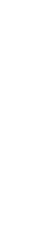}
\end{minipage}%
\begin{minipage}{.48\textwidth}
  \centering
  \includegraphics[width=0.7\linewidth]{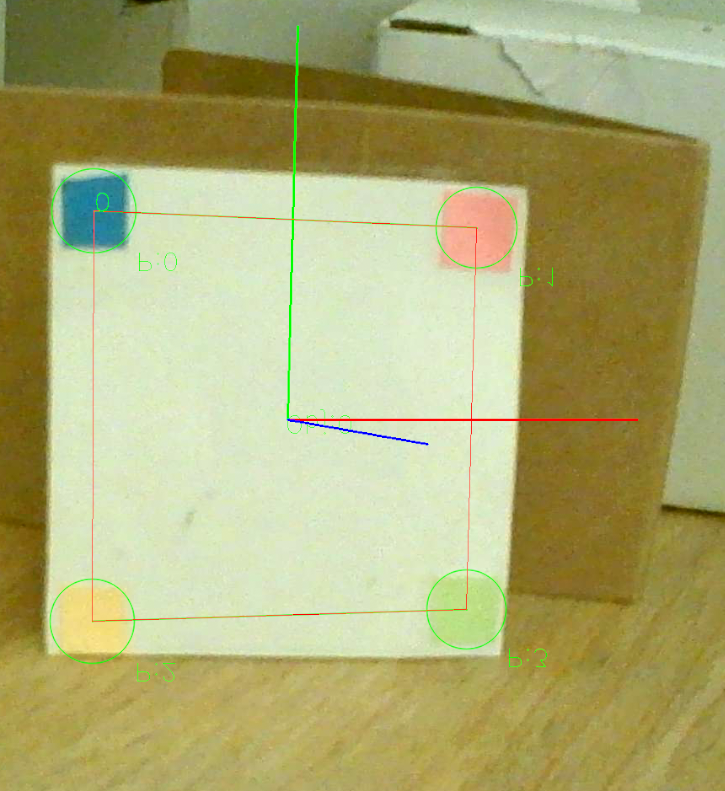}
  \captionof{figure}{A square marker detected by means of colored points. Source: own elaboration.}
  \label{fig:colorsqr}
\end{minipage}
\end{figure}

OpenCV \cite{6240859} is employed extensively for tracking and pose estimation, aided by the AprilTag \cite{AprilTag} library. AruCo and AprilTag (shown in Figure \ref{fig:arcuosheet}) systems are used in tandem to detect markers and estimate their poses using the IPPE \cite{collins2014infinitesimal} method. The Custom Colored Points method also base on IPPE, but those are recognized by the use of a custom clustering algorithm, which is faster than OpenCV's, at the cost of interference resistance.

\vspace{-12pt}

\subsection{The Colored Points Algorithm}
\begin{figure}[htbp!]
    \centering
    \includegraphics[width=\textwidth]{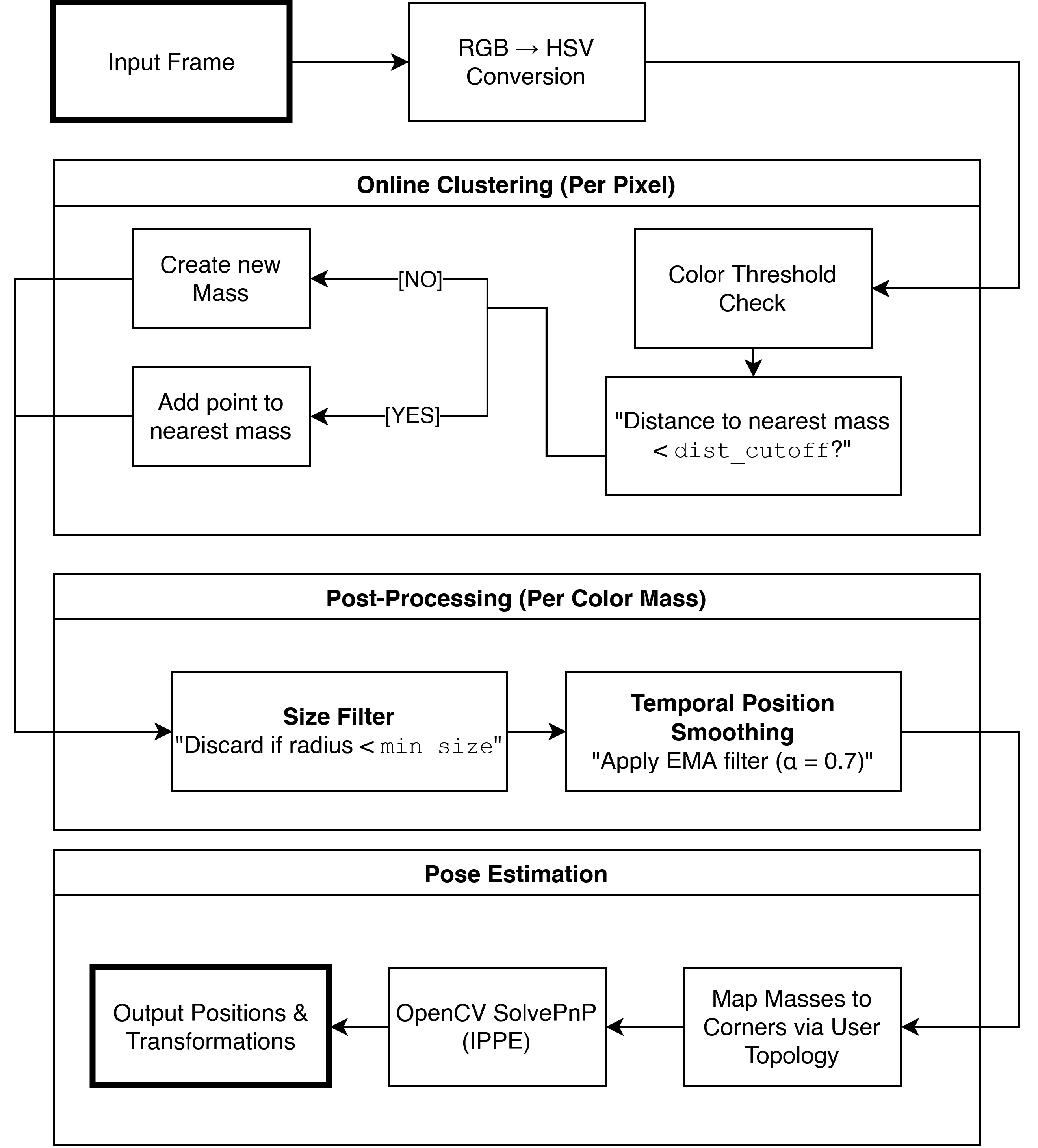}
    \caption{Block diagram of the Colored Points algorithm, illustrating the single-pass clustering and topology-based pose estimation pipeline. Source: own elaboration.}
    \label{fig:cp_flowchart}
\end{figure}

To address the limitations of standard binary markers in very low-latency or low-resolution conditions, we developed the Colored Points method. Unlike AruCo or AprilTag, which rely on high-contrast binary edges, our method utilizes distinct chromatic "islands" to define marker geometry. This allows for proper detection even when the "marker" is composed of spatially separated colored objects (e.g., LEDs or stickers), rather than a flat printed square. Figure \ref{fig:cp_flowchart} illustrates the algorithm's logic flow.

\subsubsection{Detection Pipeline.}
Prior to the core clustering logic, the pipeline applies optional pre-processing to enhance signal-to-noise ratio. To mitigate environmental interference, a \textbf{Background Removal} stage employs a static frame difference method. During setup, a reference background frame $B$ is captured. For each subsequent input frame $F_t$, the system computes the pixel-wise absolute difference $D = |F_t - B|$. This difference map is converted to grayscale and filtered against a user-defined threshold (default $\tau=50$) to generate a binary foreground mask, which is applied to the frame to isolate physical markers. \\

Following preprocessing, the algorithm operates in a single-pass online clustering mode:
\begin{enumerate}
    \item \textbf{HSV Classification:} The system iterates sequentially through the frame buffer. To ensure robustness against lighting changes, pixel values are converted to the \textbf{HSV color space}. A "Color Proximity" check is performed against pre-configured Hue, Saturation, and Value ranges.
    \item \textbf{Online Clustering:} We employ a distance-based clustering approach. Valid pixels are grouped into "Color Masses" based on Euclidean distance. If a pixel is within \texttt{dist\_cutoff} distance of an existing mass, it expands that mass's bounding box; otherwise, a new mass is initialized.
    \item \textbf{Temporal Smoothing:} To counteract sensor noise, the center coordinates of each mass are stabilized using an exponential moving average (EMA) filter ($\alpha=0.7$), blending the current position with the previous frame's result.
\end{enumerate}

\subsubsection{Pose Estimation and Topology.}
A critical distinction of our method is how it resolves geometric ambiguity. While AprilTags are self-identifying, Colored Points relies on a \textbf{user-defined topology}.
\begin{enumerate}
    \item \textbf{Object Definition:} The user defines an "Object" by specifying two logical "Lines" that connect specific Color Masses (e.g., \textit{"Mass \#0 connects to Mass \#1"}). This explicit definition assigns a fixed index to each corner (Top-Left, Top-Right, etc.).
    \item \textbf{Correspondence Solving:} The system extracts the centers of the four defined masses in the order specified by the topology. These four 2D image points are paired with a fixed 3D model (a square centered at origin, with size corresponding to the marker).
    \item \textbf{PnP Solution:} Finally, OpenCV's \texttt{solve\_pnp} calculates the rotation and translation vectors that map the 3D model to the observed 2D configuration.
\end{enumerate}

\begin{figure}
    \centering
    \includegraphics[width=0.9\textwidth]{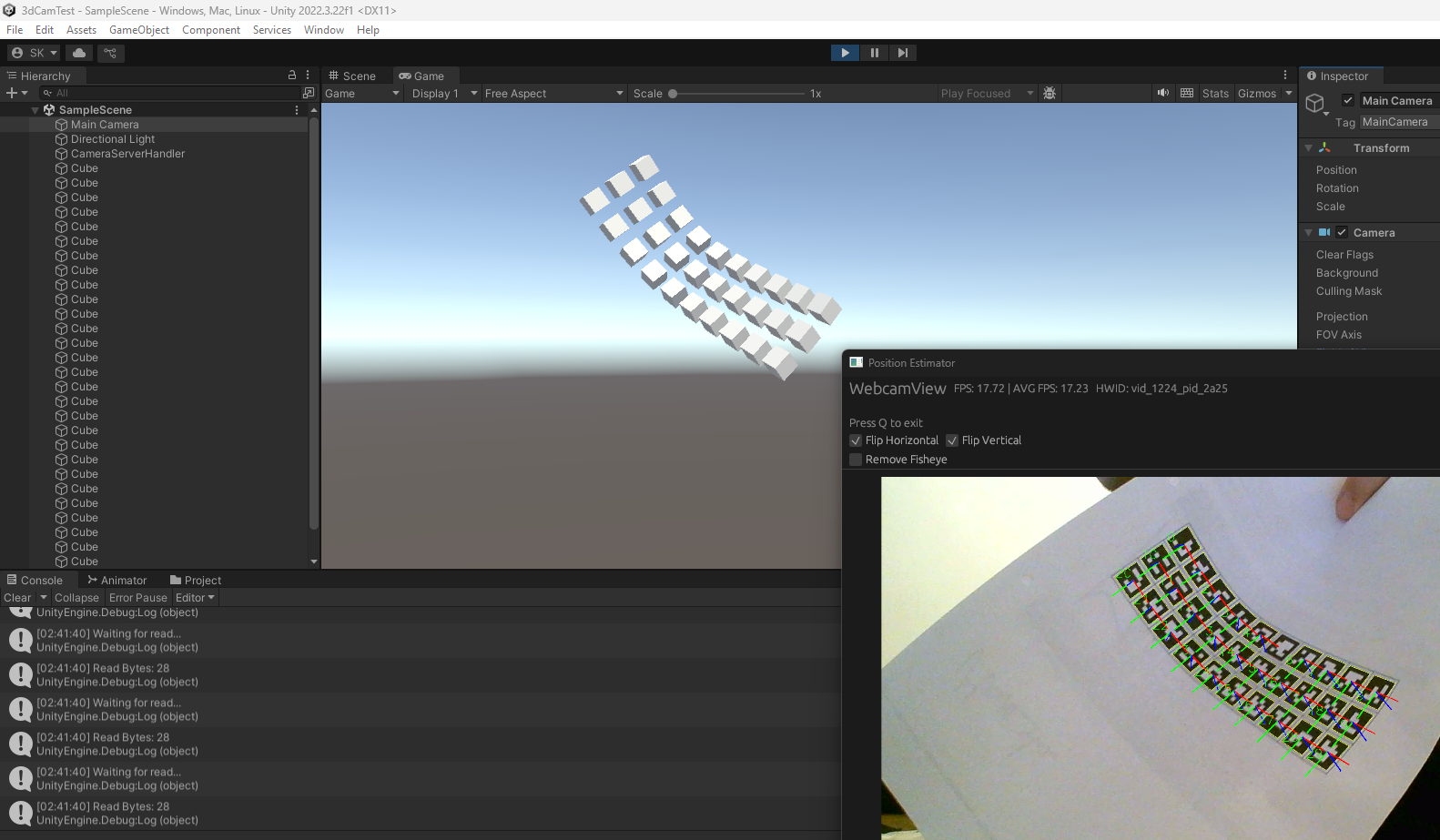}
    \caption{AprilTags detected as Unity objects. Source: own elaboration.}
    \label{fig:april-unity}
\end{figure}

The result is sent using Unix Sockets \cite{coffield1987tutorial} to the consumer program, such as Unity Engine. Figure \ref{fig:april-unity} shows the markers appearing as cubes within the engine.

\section{Results and Discussion}
\begin{figure}
    \centering
    \includegraphics[width=\textwidth]{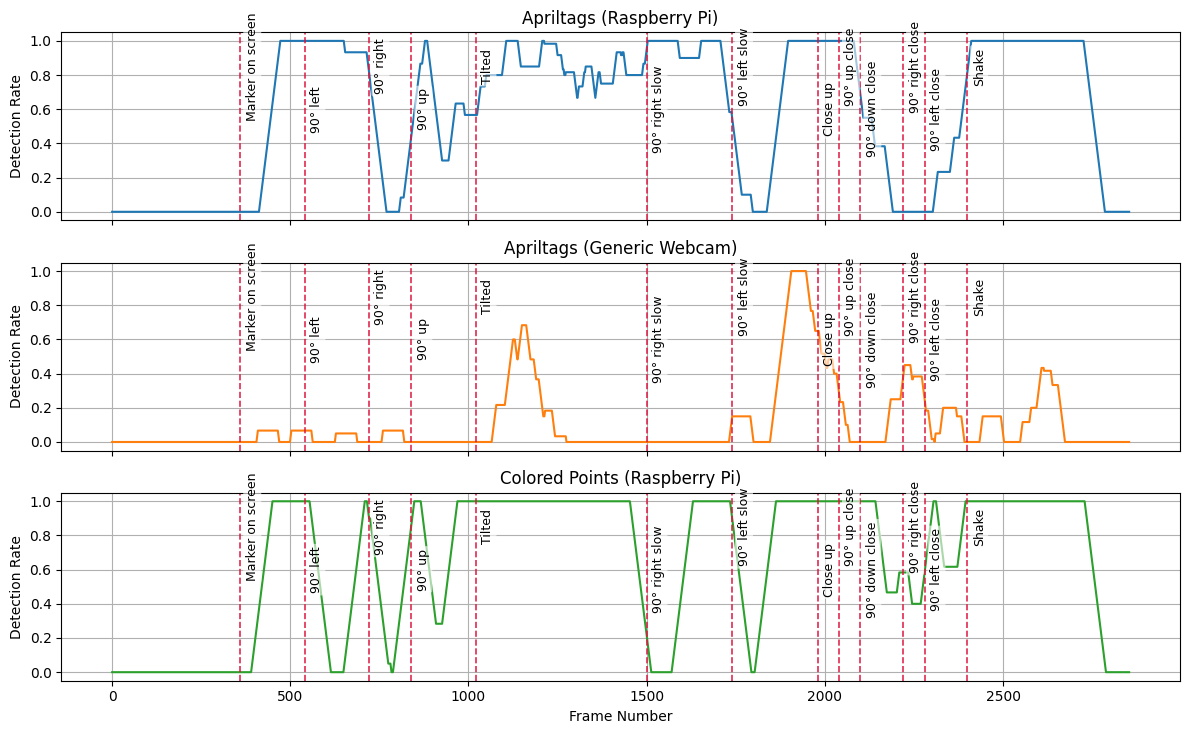}
    \caption{Timeline of detection rate. Note: Colored Points on Generic Webcam not included since they had 0\% detection rate. Source: own elaboration.}
    \label{fig:det_rate_td}
\end{figure}

The performance of the system depends on the camera quality, which significantly affects detection range and rate. We evaluated it using a Generic WebCam (2MP) and a Raspberry Pi Camera Module 3 (12MP).
To quantify stability, we defined the \textbf{Detection Rate} as a rolling average probability. Since a marker is effectively binary (Detected/Not Detected) in any single frame, the graphs in this section (see Figure \ref{fig:det_rate_td}) represent the percentage of successful detections over a sliding window of 60 frames. A value of 0.8 indicates the marker was successfully resolved in 80\% of the last 60 frames.

\begin{figure}
    \centering
    \includegraphics[width=0.6\textwidth]{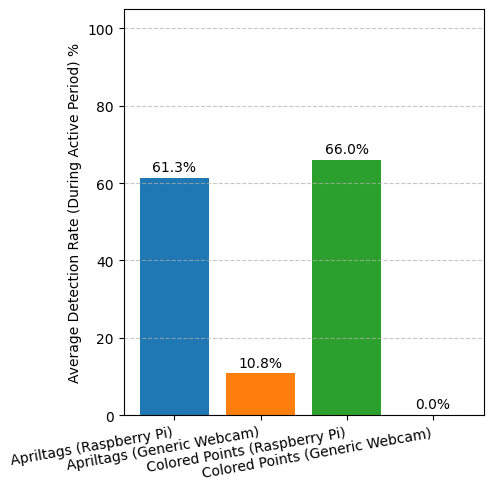}
    \caption{Average detection rate. Source: own elaboration.}
    \label{fig:det_rate_avg}
\end{figure}

\begin{figure}[htbp!]
    \centering
    \includegraphics[width=0.6\linewidth]{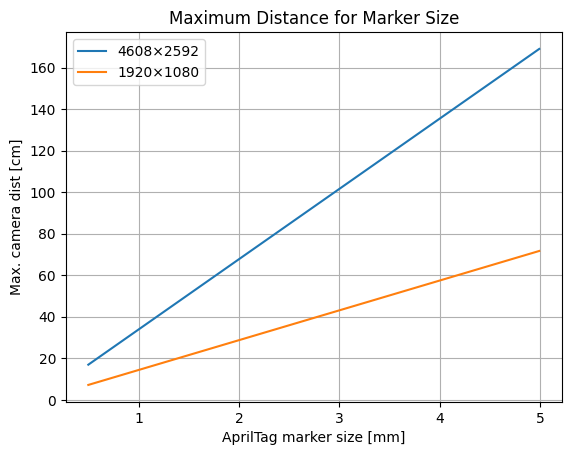}
    \caption{Maximum distance from camera for various marker sizes. Source: own elaboration.}
    \label{fig:marker_size_vs_distance}
\end{figure}

Figure \ref{fig:det_rate_avg} illustrates the average performance across different devices and algorithms, while Figure \ref{fig:marker_size_vs_distance} compares the maximum effective tracking distance, between the 12MP native resolution and the down-scaled 1080p output. 

This significance of camera resolution is best illustrated by Table \ref{tab:video_test_results}. The test was performed using parallel video of a marker, in various angles and distances, using both cameras and various detection methods. Colored Points allowed markers to be much smaller. See Table \ref{tab:marker_size_det} for details. Tests were performed by placing marked objects at various distances and noting at which distance-point the detection rate fell below 90\% of captured frames.
These empirical results closely match the theoretical limitations of the cameras, stemming from their matrix pixel size. 

\begin{table}[htbp]
    \centering
    \caption[Video test results at 1080p]{Video test results at 1080p. Source: own elaboration.}
    \begin{tabular}{|l|ll|ll|}
        \hline
        \textbf{Device} & \multicolumn{2}{l|}{\thead{\textbf{Raspberry Pi} \\ \textbf{Camera Module 3}}} & \thead{\textbf{Generic} \\ \textbf{WebCam}} & \\ \hline
        \textbf{Detection Method} & \multicolumn{1}{l|}{\textbf{AprilTag (4mm)}} & \textbf{Colored Point} & \textbf{AprilTag (4mm)} & \\  \hline
        max. normal angle hor. dev. & \multicolumn{1}{l|}{$\approx170\degree$} &  $\approx45\degree$ & N/A & \\ \hline
        max. normal angle ver. dev. & \multicolumn{1}{l|}{$\approx45\degree$} & $\approx45\degree$ & N/A & \\ \hline
        max. distance & \multicolumn{1}{l|}{$\approx46\text{cm}$} &  $\approx40\text{cm}$ & $\approx15\text{cm}$ & \\ \hline
        high movement speed det. & \multicolumn{1}{l|}{YES} & YES & NO & \\ \hline
        \% of time detected & \multicolumn{1}{l|}{73.8\%} & 79.3\% & 14.0\% & \\ \hline
    \end{tabular}
    \small{Note: Generic WebCam detection rate was too small to rate angle deviations.}
    \label{tab:video_test_results}
\end{table}

\section{Future Work}
Although the proposed system delivers a functional end-to-end pipeline, several possibilities exist for enhancing stability and robustness. 
First, while the current implementation employs basic 2D point smoothing, it lacks \textbf{pose-level continuity checks}. Specifically, the system does not currently apply heuristics to resolve the ambiguity in surface normal orientation, a phenomenon stemming from the inherent dual-solution nature of the PnP problem for planar targets \cite{collins2014infinitesimal}. Without temporal filtering on the 3D pose itself, image noise can cause the solver to oscillate between mathematically valid but physically inconsistent solutions. Future work should implement Kalman filtering or optical flow constraints to enforce orientation continuity.\\

\begin{table}[htbp]
    \centering
    \caption[Marker Sizes for stable detection at various distance thresholds at 1080p]{Marker Sizes for stable detection at various distance thresholds at 1080p. Source: own elaboration.}
    \begin{tabular}{|l|ll|ll|}
        \hline
        \textbf{Device} & \multicolumn{2}{l|}{\thead{\textbf{Raspberry Pi} \\ \textbf{Camera Module 3}}} & \multicolumn{2}{l|}{\thead{\textbf{Generic} \\ \textbf{Webcam}}} \\ 
        \hline
        \textbf{Detection Method} & \multicolumn{1}{l|}{\textbf{AprilTag}} & \multicolumn{1}{l|}{\textbf{Colored Points}} & \multicolumn{1}{l|}{\textbf{AprilTag}} & \multicolumn{1}{l|}{\textbf{Colored Points}} \\
        \hline
        stable det. at $ d\leq 25\text{cm}$ & \multicolumn{1}{l|}{$\geq2\text{mm}$} & \multicolumn{1}{l|}{$\geq1\text{mm}$} & \multicolumn{1}{l|}{$\geq7\text{mm}$} & \multicolumn{1}{l|}{$\geq4\text{mm}$} \\
        \hline
        stable det. at $ d\leq 50\text{cm} $ & \multicolumn{1}{l|}{$\geq4\text{mm}$} & \multicolumn{1}{l|}{$\geq2\text{mm}$} & \multicolumn{1}{l|}{$\geq12\text{mm}$} & \multicolumn{1}{l|}{$\geq5\text{mm}$} \\
        \hline
    \end{tabular}
    \label{tab:marker_size_det}
\end{table}

Second, the system could be extended to support a \textbf{multi-view sensor fusion framework}. Synchronizing multiple cameras would not only extend the effective angular range of markers but also resolve pose ambiguities through stereoscopic constraints and mitigate occlusion risks. 
Finally, the Colored Points method would benefit from \textbf{adaptive background subtraction} to further reduce environmental interference in dynamic lighting conditions. 

\newpage
\begin{sloppypar}
\bibliographystyle{splncs04}
\bibliography{bibliography}
\end{sloppypar}
\end{document}